\documentclass{appolb}
\usepackage{graphicx}
\usepackage{amsmath,amssymb}
\begin{document}

\title{Superconductivity in moiré transition metal dichalcogenide bilayers: comparison of two distinct theoretical approaches%
}
\author{Waseem Akbar and Michał Zegrodnik
\address{Academic Centre for Materials and Nanotechnology, AGH University of Krakow, al. A. Mickiewicza 30, 30-059 Krakow, Poland}
%\\[3mm]
%{Third Author % of different affiliation
%\address{affiliation}
%}
%\\[3mm]
%the Name(s) of other Author(s)
%\address{affiliation}
}
\maketitle
\begin{abstract}
Superconductivity has recently been observed in moir\'e transition-metal dichalcogenide bilayers. Here, we investigate the superconducting state in twisted WSe$_2$ using two complementary theoretical approaches. The first is based on the negative $U$-Hubbard model and represents a relatively conventional pairing scenario, in which strong electron–electron repulsion does not directly affect the paired state and an isotropic $s$-$wave$ gap emerges. The second approach employs the $t$-$J$-$U$ model, allowing for unconventional gap symmetries and incorporating strong correlation effects via substantial renormalization induced by Coulomb repulsion. We compare the key properties of the superconducting states obtained within these two frameworks and discuss their implications in light of available experimental observations.
\end{abstract}

\section{Introduction}

Moir\'e transition metal dichalcogenide (TMD) bilayers have emerged as a highly tunable platform for exploring correlated electron physics in two dimensions. When two monolayers are stacked with a small twist angle or lattice mismatch, long-wavelength moiré superlattices form, leading to flat electronic bands. These systems combine strong spin–orbit coupling, valley degrees of freedom, and reduced dielectric screening, giving rise to a rich variety of interaction-driven ground states including correlated insulators\cite{exp_insul_1,exp_insul_2,Zhao2023}, topological phases\cite{exp_top_1,exp_top_2,exp_top_3,mw_exp_1,b_field,chern_1,chern_2}, generalized Wigner crystals\cite{exp_wig_2,Emma2020} and tunable excitonic complexes\cite{Kausik_2022,Zheng_2023,Wang2025}. Moreover, recent experiments on twisted bilayer WSe$_2$ and MoTe$_2$ have revealed the appearance of the superconducting phase (SC)\cite{Wang2020,Yiyu2025,Yinji2025,xia2025_arxiv,Dean2025,Li_2025}. It should be noted that TMD moir\'e bilayers are characterized by significant advantages in comparison to the canonical strongly correlated electron systems such as cuprates, in which superconductivity was also identified many years ago\cite{bednorz1986possible,Spalek2022}. Namely, by varying the twist angle between the layers, one directly controls the bandwidth and thus the strength of electronic correlations, enabling continuous tuning from weak- to strong-coupling regimes. Independent electrostatic control of the band filling through gate voltages further allows superconductivity to be probed across distinct carrier-density regimes. In addition, the strong spin–orbit coupling intrinsic to WSe$_2$ can be modified via an applied displacement field, altering the spin–valley locking and possibly modifying the available pairing channels. Together, these tuning knobs make twisted bilayer WSe$_2$ a uniquely flexible platform for investigating unconventional superconductivity in two dimensions, where both the interaction scale and internal electronic structure can be engineered in an $in$ $situ$ manner. 

According to the experimental data gathered so far, the superconductivity in twisted WSe$_2$ appears close to the correlated insulator, which lies at half-filling. The paired state shows dome-like behavior as a function of electronic concentration with the maximal critical temperature below $1\;$K. For small twist angles ($\sim2^{\circ}$) the SC state stability appears close to zero displacement fields, while for relatively larger angles ($\sim5^{\circ}$) some non-zero displacement field is necessary to induce the pairing\cite{xia2025_arxiv,Li_2025}. Also, depending on the twist angle and the value of displacement field, both one and two superconducting domes can be observed with changing band filling. 

So far the pairing mechanism and the complete theoretical description of the fundamental features of the superconducting phase in tWSe$_2$ has not been established. Most of the theoretical analysis emphasize the Coulomb-interaction driven pairing both within strong and weak-to moderate interaction regime which leads to an anisotropic gap in $\mathbf{k}$-space and singlet-triplet mixing\cite{Zegrodnik2023,Waseem2024,Millis2024_arxiv,Tuo2024arxiv,Klebl2022,Millis2025,Fengcheng2025,Millis2023,Chen2023,zegrodnik2025}. However, pure spin-singlet channel has also been considered\cite{Chowdhury2025,Belanger2022_cDMFT_SC} and a more standard phonon mediated pairing still remains a possibility and has been discussed in Ref. \cite{DasSarma20225prb}.

The main aim of this paper is to confront two distinct approaches to the description of superconductivity as applied to the study of tWSe$_2$ and to discuss them in the context of available experimental data. Within the first scenario, we apply the so-called negative-$U$ Hubbard model which was originally proposed by Robaszkiewicz et al. and analyzed in the context of superconducting state and charge ordering\cite{Micnas1981,Micnas1990}. Here we combined it with an effective description of a single moi\'e band of tWSe$_2$ with the Ising-type spin orbit coupling taken into account by the spin dependent complex hoppings. This constitutes a relatively conventional approach and leads in a straightforward manner to an isotropic $s$-$wave$ type of pairing. Such description can be considered as compatible with momentum independent pairing potential similar to the one considered in the standard BCS theory but without limiting the pairing to a narrow range of energies around the Fermi level. 

In the second case, we revisit our recent considerations in which we employ the $t$-$J$-$U$ model 
that refers to electron pairing induced by kinetic exchange leading to an unconventional form of the superconducting gap realizing a mixed $d+id$/$p- ip$ symmetry, which is due to the presence of Ising type spin-orbit coupling in tWSe$_2$\cite{Zegrodnik2017,zegrodnik2025}. It should be noted that the $t$-$J$-$U$ model has been extensively applied to superconducting cuprates already some time ago\cite{Zegrodnik_2017,Spalek2022} and it can be considered as evolving from the canonical Hubbard\cite{Hubbard_1963,Bulut01112002} and $t$-$J$\cite{Spalek_1977,Ogata2008} model based theories within the paradigm of strong electronic correlations. Here, we use the Gutzwiller approximation method in order to analyze the influence of the correlation induced renormalization effects and their influence of the resulting physical picture.

\section{Model and Method}
\begin{figure}[htb]
\centerline{%
\includegraphics[width=12.0cm]{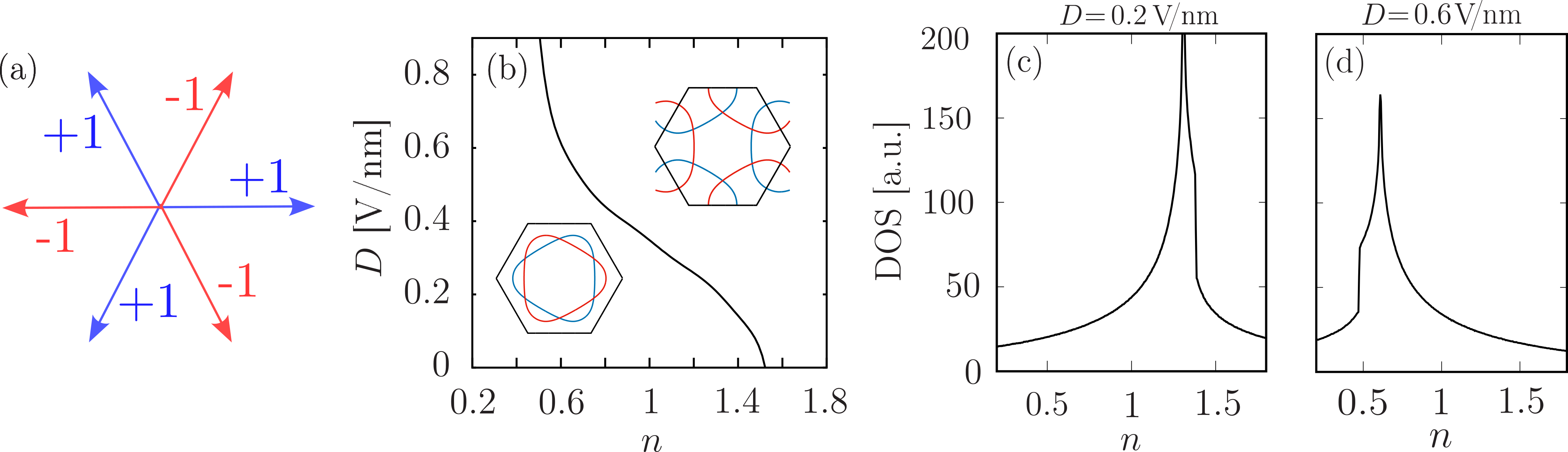}}
\caption{(a) The values of the direction dependent factor $\nu_{ij}=\pm1$ corresponding to the nearest neighbor hoppings on a triangular lattice which determines the sign of the complex phase appearing in Eq. (\ref{eq:Ham_0}). (b) The evolution of the Van Hove singularity in the ($n$, $D$)-plane. Note that the resulting curve divides the diagram into two regimes with closed (left-bottom) and opened (right-upper) spin-split Fermi surfaces. (c) and (d) Density of states at the Fermi level as a function of band filling for two selected values of the displacement field, $D$.}
\label{Fig:fig0}
\end{figure}

To describe the single moir\'e band of twisted WSe$_2$ with Ising type spin orbit coupling in an effective manner, we consider the following single-particle Hamiltonian
\begin{equation}
    \begin{split}
    \hat{H}_0&=t\sum_{\langle ij\rangle\sigma}e^{i\sigma\nu_{ij}\phi}\;\hat{c}^{\dagger}_{i\sigma}\hat{c}_{j\sigma}, \\
    \end{split}
    \label{eq:Ham_0}
\end{equation}
where $\hat{c}^{\dagger}_{i\sigma}$ and $\hat{c}_{i\sigma}$ are the creation and annihilation operators for the electron with spin $\sigma$ at the site $i$ while $\langle i,j\rangle$ corresponds to the nearest neighbor lattice sites at the triangular lattice. As one can see, the hoppings are both spin- and direction-dependent with $\sigma=1$ ($\sigma=-1$) corresponding to spin up (spin down) and $\nu_{ij}=\pm 1$ depending on the hopping direction [cf. Fig. \ref{Fig:fig0} (a)]. 

It should be noted that within such an effective single band picture, the hopping amplitude $t$ and the phase factor $\phi$ depend on the displacement field ($D$) as shown in Refs. \cite{Wang2020,Haining2020}. Here we take the values provided in \cite{Wang2020}. By changing the displacement field, one can enhance the strength of the Ising type spin orbit coupling and influence the profile of the density of states in the system. In Fig. \ref{Fig:fig0} (b) we show the resulting evolution of the Van Hove singularity in the ($n$, $D$)-plane, where $n=\sum_{\sigma}\langle \hat n_{i\sigma}\rangle$. The presented curve divides the diagram into two regimes with closed (left-bottom) and opened (right-upper) spin-split Fermi surfaces. For the sake of clarity in Fig. \ref{Fig:fig0} (c) and (d) we provide the density of states at the Fermi level as a function of band filling for two selected displacement fields, where the shift of the Van Hove singularity from higher to lower band fillings is visualized with increasing $D$.

In this study, the single particle part provided above is supplemented with an interaction part in order to analyze the fundamental features of the superconducting state. Therefore the resulting Hamiltonian has the general form
\begin{equation}
    \hat{H}=\hat{H}_0+\hat{H}_I,
    \label{eq:Ham}
\end{equation}
where $\hat{H}_I$ constitutes the interaction part. In the following, we describe the two theoretical approaches which differ by the form of the interaction term and correspond to the negative $U$-Hubbard model in connection with the Hartree-Fock method as well as the $t$-$J$-$U$ model with the application of the Gutzwiller approximation method.

\subsection{Negative U-Hubbard model}
First we consider the interaction part of our model in the form of the negative $U$-Hubbard term
\begin{equation}
    \hat{H}_I=U\sum_{i}\hat{n}_{i\uparrow}\hat{n}_{i\downarrow},
    \label{eq:Ham_I_negative_U}
\end{equation}
where $\hat{n}_{i\sigma}=\hat{c}_{i\sigma}^{\dagger}\hat{c}_{i\sigma}$ and $U<0$ leading to an attractive onsite interaction which results in superconducting pairing in a straightforward manner as shown below. After the application of the Hartree-Fock approximation with the inclusion of the anomalous expectation values, the interaction term takes the form
\begin{equation}
    \hat{H}_{I}=\sum_i\big(\tilde{\Delta}^*\hat{c}_{i\uparrow}\hat{c}_{i\downarrow}+\tilde{\Delta}\hat{c}^{\dagger}_{i\downarrow}\hat{c}^{\dagger}_{i\uparrow}\big)+U\frac{n}{2}\sum_{i\sigma}\hat{n}_{i\sigma}-\frac{N}{U}|\tilde{\Delta}|^2-NU\frac{n^2}{4},
    \label{eq:Ham_I_negative_U_HF}
\end{equation}
 where $N$ is the number of lattice sites and we have introduced the onsite pairing amplitude $\tilde{\Delta}_i=U\langle\hat{c}_{i\downarrow}\hat{c}_{i\uparrow} \rangle$. Also, since we consider a homogeneous, non-magnetic state, we take $\tilde{\Delta}_i\equiv\tilde{\Delta}$ and $\langle \hat{n}_{i\sigma}\rangle\equiv n/2$. For the sake of convenience, we additionally determine the pure anomalous expectation value $\Delta=\tilde{\Delta}/U$, which is going to be useful further on when comparing the results coming from the two considered here approaches. Moreover, the second summation in the above equation can be dropped since it only shifts the energy scale of the considered band and therefore in the considered single band case it does not have any influence on the physics of the model.

After the transformation to reciprocal space we can write the full Hamiltonian in a compact form
\begin{equation}
\begin{split}
\hat{H}&=\frac{1}{2}\sum_{\mathbf{k}\sigma}\begin{array}{c}
 \left(\hat{c}^{\dagger}_{\mathbf{k}\sigma}\;\; \hat{c}_{-\mathbf{k}\bar{\sigma}} \right)\\
 \\
\end{array}\left(\begin{array}{cc}
 \epsilon_{\mathbf{k}\sigma}-\mu & \sigma\tilde{\Delta}\\
\sigma\tilde{\Delta}^{*} & -\epsilon_{-\mathbf{k}\bar{\sigma}}+\mu \\
\end{array} \right)\left(\begin{array}{c}
 \hat{c}_{\mathbf{k}\sigma}\\
 \hat{c}^{\dagger}_{-\mathbf{k}\bar{\sigma}}\\
\end{array} \right)\;\\
&+\sum_{\mathbf{k}\sigma}(\epsilon_{\mathbf{k}\sigma}-\mu)-\frac{N}{U}|\tilde{\Delta}|^2-NU\frac{n^2}{4},
\end{split}
\label{eq:H_negative_U_k_space}
\end{equation}
where $\bar{\sigma}=-\sigma$ and we have supplemented the model with the chemical potential contribution ($-\mu\sum_i\hat{n}_{i\sigma}$). As one can see, onsite pairing amplitudes $\tilde{\Delta}$ correspond to coupling between the electron- and hole-like dispersion relations which leads to gap opening in the resulting band structure. The standard Bogolubov-de Gennes approach can be applied to the Hamiltonian given by Eq. (\ref{eq:H_negative_U_k_space}) in order to derive the self consistent equations for the superconducting gap and chemical potential, which than can be solved numerically.

\subsection{The $t$-$J$-$U$ model}
As the second scenario considered here we revisit the $t$-$J$-$U$ model as applied to the single band effective description of the twisted WSe$_2$\cite{Waseem2024,zegrodnik2025}. In this case the interaction part of the model is taken in the form
\begin{equation}
\begin{split}
 \hat{H}_I=& J\sideset{}{'}\sum_{\langle ij\rangle}\bigg(\hat{S}^z_i\hat{S}^z_j +\cos{(2\nu_{ij}\phi)}\sum_{\alpha=x,y}\hat{S}^{\alpha}_i\hat{S}^{\alpha}_j\\
 &+\sin{(2\nu_{ij}\phi)}(\mathbf{\hat{S}}_i\times\mathbf{\hat{S}}_j)\cdot \hat{z} \bigg)+U\sum_{i} \hat{n}_{i\uparrow} \hat{n}_{i\downarrow},
 \label{eq:Hamiltonian_start}
 \end{split}
\end{equation}
where $\mathbf{\hat{S}}_i$ is the spin-$\frac{1}{2}$ operator while $J>0$ and $U>0$ correspond to intersite kinetic exchange and onsite Coulomb repulsion, respectively. The presented form of the kinetic exchange term can be obtained when deriving the $t$-$J$ model form the Hubbard model description of a single moir\'e band of tWSe$_2$. The considered $t$-$J$-$U$ Hamiltonian can be viewed as an effective framework appropriate for regimes where $U$ is large but finite, and double occupancies remains small yet nonvanishing. Under these conditions, the Hubbard $U$ term must be retained in the Hamiltonian, while the kinetic spin–spin exchange interaction already starts to develop. The $t$-$J$-$U$ model is therefore used to explicitly capture both contributions. It should be noted that, here the kinetic exchange term differs from the one obtained originally by Spa\l ek et al\cite{Spalek_1977} due to the appearance of the additional Dzyaloshinskii-Moriya-type of term. Such a form is a straightforward result of the Ising type spin orbit coupling which is characteristic to tWSe$_2$ and is taken into account via spin and direction dependent complex hoppings. 

In order to take into account the electron-electron correlations within such approach we consider the Gutzwiller variational wave function in the following form
\begin{equation}
    |\Psi_G\rangle=\prod_i\sum_{\Gamma}\lambda_{i,\Gamma}|\Gamma\rangle_{i\;i}\langle\Gamma|\Psi_0\rangle,
    \label{eq:correlated_state}
\end{equation}
where $|\Psi_0\rangle$ is the uncorrelated (mean-field) state and $i$ runs over all the lattice sites, $|\Gamma\rangle$ corresponds to the local basis $|\Gamma\rangle\in \{ |\emptyset\rangle,\;|\uparrow\rangle,\;|\downarrow\rangle,\;|\uparrow\downarrow\rangle \}$, and $\lambda_{i,\Gamma}$ are the variational parameters. Similarly as in the previous subsection, also here we consider a homogeneous, non-magnetic state, thus  $\lambda_{i,\Gamma}\equiv\lambda_{\Gamma}$ and $\lambda_{\sigma}\equiv\lambda_s$. As shown by Bünemann et al., one can impose
an additional condition to the correlation operator, in order to reduce the complexity of the numerical calculations\;\cite{Bunemann_2012}. Within the infinite dimensions approximation (Gutzwiller approximation), the expectation value of our Hamiltonian in the $|\Psi_G\rangle$ state takes the form 
\begin{equation}
\begin{split}
    \langle\hat{H}\rangle_G&=\sum_{ij}q^2t_{ij\sigma}\langle\hat{c}^{\dagger}_{i\sigma}\hat{c}_{j\sigma}\rangle_0+\lambda^4_s\;J\sideset{}{'}\sum_{\langle ij\rangle}\bigg(\frac{1}{2}\;\sum_{\sigma}\;e^{i\sigma2\phi_{ij}}\langle\hat{c}^{\dagger}_{i\sigma}\hat{c}_{i\bar{\sigma}}\hat{c}^{\dagger}_{j\bar{\sigma}}\hat{c}_{j\sigma} \rangle_0\\
    &+\frac{1}{4}\;\sum_{\sigma\sigma'}\sigma\sigma'\langle\hat{n}^{HF}_{i\sigma}\hat{n}^{HF}_{j\sigma'}\rangle_0\bigg)+\lambda_{\uparrow\downarrow}^2U \sum_{i}\langle \hat{n}_{i\uparrow}\hat{n}_{i\downarrow}\rangle_0,
\end{split}
\label{eq:hamiltonian_expectation_value}
\end{equation}
where $q=\lambda_s\big(\lambda_{\uparrow\downarrow} n/2+\lambda_{\emptyset}(1-n/2)\big)$ and $\langle\hat{o}\rangle_0$ stands for the expectation value of the $\hat{o}$ operator in the state $|\psi_0\rangle$. Using the standard Wick’s theorem, all four-operator expectation values appearing on the right-hand side of the above equation can be decomposed. As a result, $\langle\hat{H}\rangle_G$ can be written as a function of $n$, the electron hopping [$ P_{ij\sigma}=\langle\hat{c}^{\dagger}_{i\sigma}\hat{c}_{j\sigma}\rangle_0$] and Cooper pairing mean-field parameters [$S_{ij}^{\sigma\sigma'}=\langle\hat{c}_{i\sigma}\hat{c}_{j\sigma'}\rangle_0$]. In order to determine the values of those mean fields we apply the Effective Hamiltonian Scheme\cite{Kaczmarczyk2015}. Within such approach the minimization condition of the ground state energy, $\langle\hat{H}\rangle_G$, leads to an effective Hamiltonian which can be treated within the Bogilubov-de Gennes formalism as shown in some more detail in Refs. \cite{Waseem2024,zegrodnik2025}.

It should be noted that the anomalous expectation values in the correlated state can be extracted from the non-correlated counterparts with the use of the following relation: $\Delta_{ij}^{\sigma\sigma'}=\langle\hat{c}_{i\sigma}\hat{c}_{j\sigma'} \rangle_G=q^2\langle\hat{c}_{i\sigma}\hat{c}_{j\sigma'} \rangle_0$. After determining the correlated pairing amplitudes to all nearest neighbors one can extract the so-called symmetry resolved gap amplitudes corresponding to different pairing symmetries, which are going to be analyzed in the following Section.

\section{Results}
\begin{figure}[htb]
\centerline{%
\includegraphics[width=12.0cm]{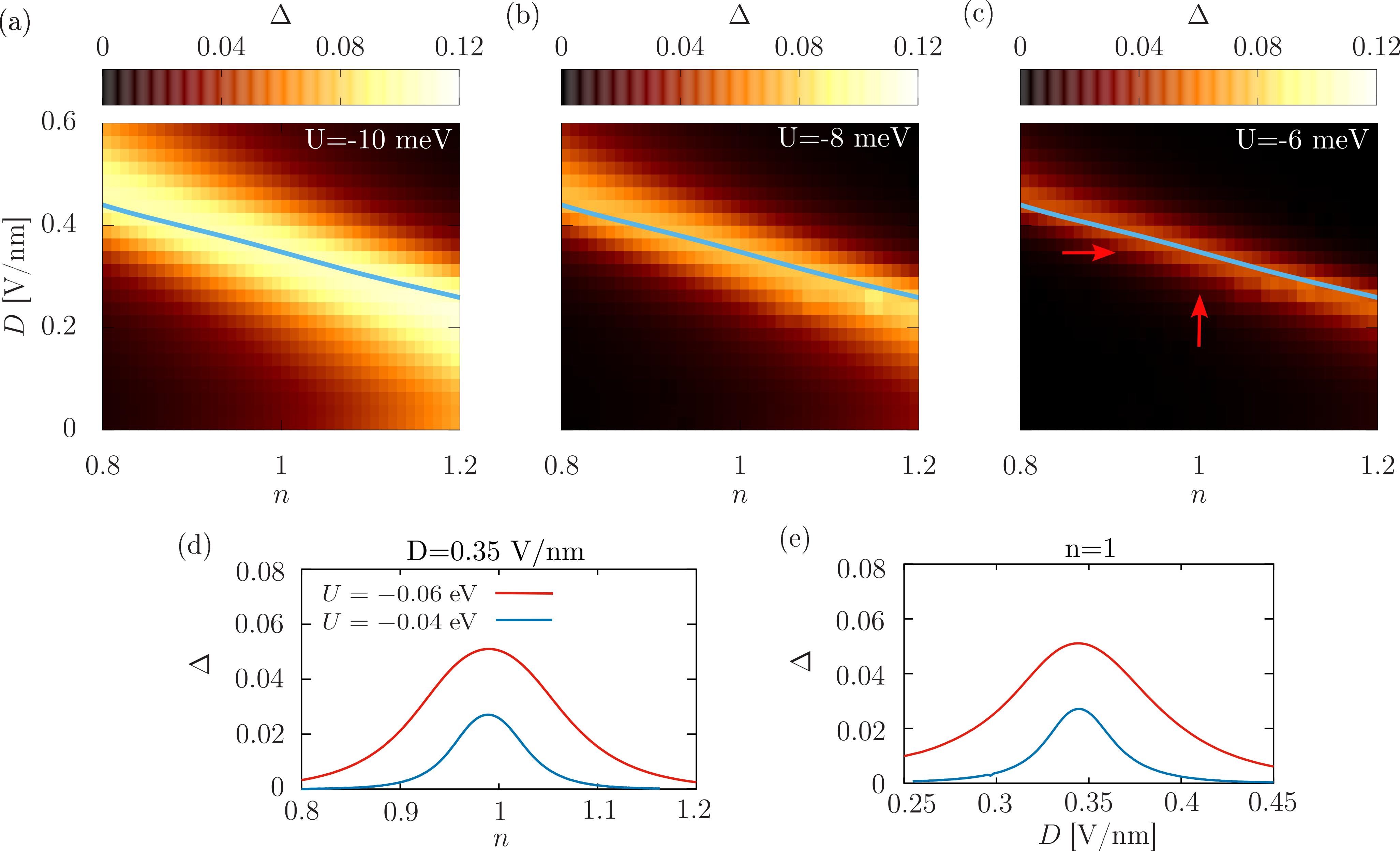}}
\caption{Superconducting gap amplitude as a function of displacement field ($D)$ and band filling $n$ resulting from the nagative $U$-Hubbard model for three selected values of the $U$ parameter. The blue solid line marks the evolution of the Van Hove singularity across the phase diagram. In (d) and (e) we show the superconducting gap amplitude as a function of band filling and displacement field, respectively, for two selected values of $U$ along the cuts represented by red arrows in (c).}
\label{fig:Dn_dep_negative_U}
\end{figure}

At first we analyze the results obtained from the negative $U$-Hubbard model as applied to the tWSe$_2$. As mentioned in the introduction from the experimental perspective the considered twisted bilayer structure is placed in between two electrodes which allows to change the band filling ($n$) as well as the bias voltage across the bilayer ($D$) in an $in$ $situ$ manner. In order to relate our theoretical results with the available experimental data we have calculated the amplitude of the superconducting gap as a function of both $n$ and $D$ which we show in Fig. \ref{fig:Dn_dep_negative_U} for three selected values of $U$. For relatively large negative $U$ the isotropic $s$-$wave$ SC state appears to be stable in a significant area of the phase diagram. Since $U$ plays the role of the pairing strength, while decreasing it, the superconductivity is becoming less and less robust. In particular, for $U=-6\;$meV the SC state survives only in close proximity of the Van Hove singularity line which is marked in blue. For such a case, the high values of density of states generated by the Van Hove singularity are necessary to stabilize superconductivity. For completeness, in Fig. \ref{fig:Dn_dep_negative_U} (d) and (e) we show the SC gap as a function of $n$ and $D$ separately along the cuts represented by the red arrows in Fig. \ref{fig:Dn_dep_negative_U} (c). Note, that the visible dome-like behavior of the gap amplitude $\Delta$ with increasing $n$ and $D$ is in qualitative agreement with the experimental data\cite{Yinji2025,Yiyu2025}. This is due to a trivial effect related with the enhancement of the SC pairing by large values of the density of states, which correspond to $n\approx1$ and $D\approx0.35\;$V/nm. Nevertheless, the calculated phase diagram in Fig. \ref{fig:Dn_dep_negative_U} does not reproduce the one observed in experiments. Since in the latter case the appearance of the paired state is reported only close to half-filling, while in the former, superconductivity stays stable around the Van Hove singularity line even away from the $n=1$ situation.

\begin{figure}[htb]
\centerline{%
\includegraphics[width=12.0cm]{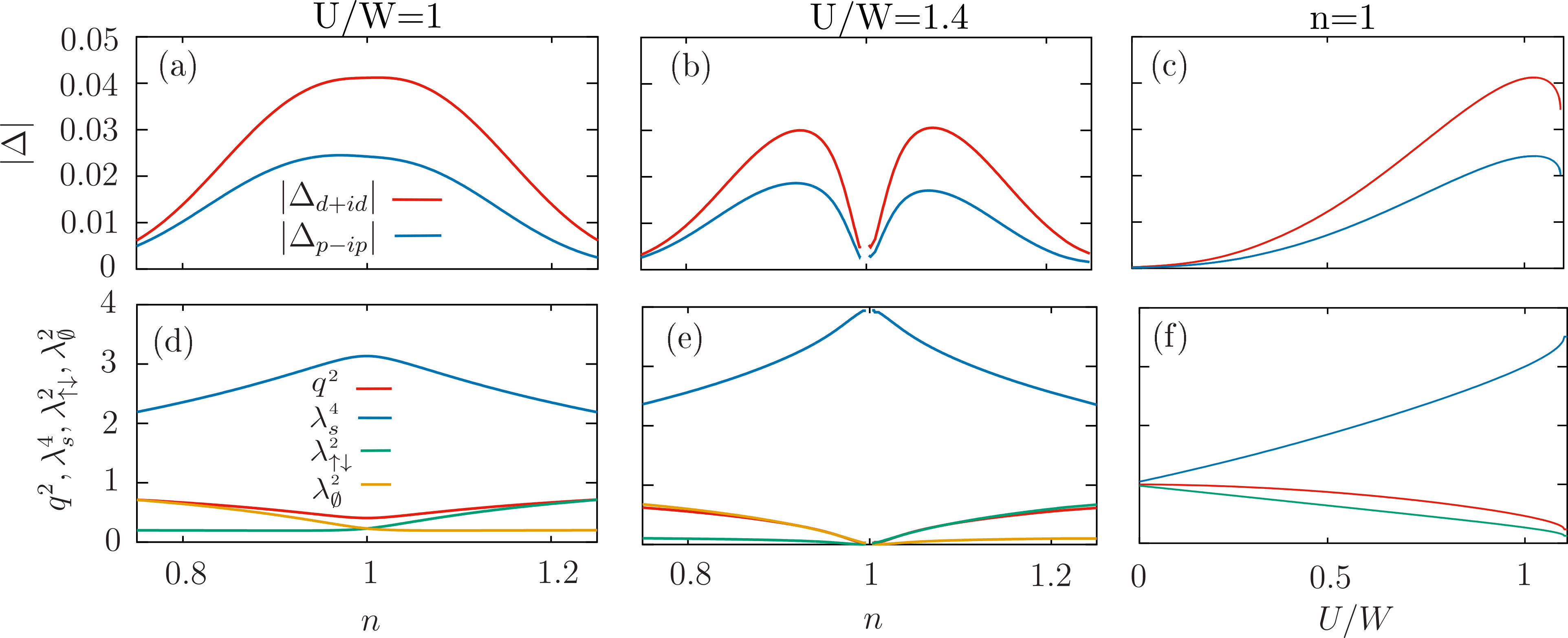}}
\caption{The gap amplitudes for the obtained mixed singlet-triplet paired state as a function of band filling (a,b) and onsite Coulomb repulsion (c) for one selected value of displacement field $D=0.35\;$V/nm. Note that $W\approx90\;$meV corresponds to the bare band width. In (d-f) we show the corresponding renormalization factors obtained within the Gutzwiller approximation [Eq. (\ref{eq:hamiltonian_expectation_value})].}
\label{fig:ndep_Udep_t_J_U_model}
\end{figure}

Next, we analyze the results obtained within the $t$-$J$-$U$ model in connection with the Gutzwiller approximation method. The first difference with respect to the previous considerations stems from the fact that here the pairing mechanism has an intersite character, meaning that the resulting SC gap is $\mathbf{k}$-dependent and can realize various symmetries. In the case considered, the stable paired state corresponds to a mixed $d+id$ (singlet) and $p-ip$ (triplet) channel and is topologically non-trivial with Chern number $C=\pm2,\pm4$, depending on the particular set of model parameters\cite{Waseem2024}. In Fig. \ref{fig:ndep_Udep_t_J_U_model} (a) and (b) we show the calculated singlet and triplet gap amplitudes as a function of band filling for the case of moderate and strong correlations, respectively. As one can see in the former case, also a single SC dome appears similarly as for the negative $U$-Hubbard model, while in the latter a two-dome structure appears with the SC state being suppressed at half-filling due to the presence of the insulating state. An important difference between this framework and the previous one is that here the strong onsite Coulomb repulsion generates renormalization of subsequent terms of the Hamiltonian [cf. Eq. (\ref{eq:hamiltonian_expectation_value})]. The values of all the renormalization parameters are provided in Figs. \ref{fig:ndep_Udep_t_J_U_model} (d-f). As can be seen, renormalization is the strongest near half-filling and increases with increasing Coulomb repulsion $U$. It should be noted that the parameter $\lambda_s^4>1$ renormalizes the kintic exchange term that is responsible for pairing, meaning that the correlations work in favor of the SC state stabilization. Nevertheless, for $U/W>1$ the SC gap is largely suppressed at half filling due to $q^2\approx0$ (since $\Delta^{\sigma\sigma'}_{ij}=q^2\langle\hat{c}_{i\sigma}\hat{c}_{j\sigma'}\rangle$). Therefore, the optimal situation for stabilization of the SC state is close to half-filling but in the moderately correlated regime where $U/W\lesssim 1$. For comparison we have carried out calculations with the use of the Hartree-Fock method for $U/W=1$ by setting all the renormalization parameters to $1$, which leads to the resulting SC gap to be few orders of magnitude smaller than the one shown in Fig. \ref{fig:ndep_Udep_t_J_U_model}(a). It should be noted that the effect of Van Hove singularity discussed earlier is going to be operative also here. However, in the moderately correlated regime the half-filled case should create favorable conditions for the appearance of the superconducting state possibly making it especially stable in the close proximity of the $n=1$ case. This is in contrast with the result presented in Fig. \ref{fig:Dn_dep_negative_U} where the value of the SC gap remained unchanged also away from the half-filling provided that $D$ is adjusted to make the Van Hove singularity stay at the Fermi level. As we have shown very recently, after supplementing the considered $t$-$J$-$U$ model with a nearest neighbor Coulomb repulsion term, which additionally suppresses the pairing in a large area of the phase diagram\cite{zegrodnik2025}, the SC state survives only at the crossing of the Van Hove singularity with the half-filling, where the two effects (renormalization and Van Hove singularity) can work simultaneously in favor of the pairing leading to a phase diagram that agrees qualitatively with the experimental data.

\section{Conclusions}
We have compared two distinct approaches to the description of the superconducting phase in a single moir\'e band of twisted WSe$_2$. In both cases, high values of the density of states stabilize superconductivity which creates a tendency for the SC state stability regime to follow the Van Hove singularity line at the ($n$,$D$)-phase diagram. For the case of negative $U$-Hubbard model the paired state realizes a trivial $s$-$wave$ symmetry of the gap and it can become stable even away from the half-filled case provided the displacement field is adjusted so as the high density of states-condition is fulfilled. Such behavior is in contrast with the experimental picture, where the SC state appears as stable only close to the half-filled scenario. 

For the case of the $t$-$J$-$U$ model calculated with the use of the Gutzwiller approximation method, the pairing appears due to intersite kinetic exchange leading to an exotic $d+id/p-ip$ mixed symmetry state. Moreover, the renormalization effects resulting from a significant onsite Coulomb repulsion can work in favor of the paired state close to half-filling in the moderately correlated state when $U\lesssim W$, for which the Mott insulator is still not well developed. This effect can cause the superconducting pairing to be specifically stable close to half-filling. As shown in more detail in Ref. \cite{zegrodnik2025}, after the inclusion of the intersite Coulomb repulsion, which suppresses the stability of the superconducting state in a large area of the phase diagram, one obtains a situation in which the SC state survives only at the crossing between the Van Hove singularity line and half-filling at the ($n$,$D$) phase diagram. In such a case, the effects of Van Hove singularity as well as correlation induced renormalization can work simultaneously stabilizing the paired phase only in a small area of the phase diagram in qualitative agreement with the experimental observations.     

\section{Acknowledgment}
This work is dedicated to Professor J\'ozef Spa\l ek in recognition of his outstanding contributions to the field of correlated electron systems, as well as his exceptional mentorship. I sincerely thank him for his guidance, support, and the many lessons that have influenced my scientific career (MZ).

This research was supported by the National Science Center, Poland (NCN) according to decision 2021/42/E/ST3/00128.

%uncomment the following lines to place a figure

\bibliography{sample}

\end{document}